\documentclass[final,12pt]{elsarticle}%

\usepackage{hyperref}
\usepackage{lineno}
\usepackage{graphicx}%
\usepackage{multirow}%
\usepackage{amsmath,amssymb,amsfonts}%
\usepackage{amsthm}%
\usepackage{mathrsfs}%
\usepackage[title]{appendix}%
\usepackage{xcolor}%
\usepackage{textcomp}%
\usepackage{manyfoot}%
\usepackage{booktabs}%
\usepackage{algorithm}%
\usepackage{algorithmicx}%
\usepackage{algpseudocode}%
\usepackage{listings}%
\usepackage{tikz}
\usetikzlibrary {perspective, patterns.meta, patterns}
\usepackage{pgfplots}
\pgfplotsset{compat=newest} 
\usepgfplotslibrary{units} 
\usepackage{siunitx}
\usepackage{url}

\usepackage{soul}

\newcommand{\revision}[1]{{#1}}

\begin{document}

\begin{frontmatter}
\title{Poromechanical modelling of the time-dependent response of \textit{in vivo} human skin during extension}


\author[1,2,3]{Thomas Lavigne} 

\author[1]{Stéphane Urcun}

\author[5]{Emmanuelle Jacquet}

\author[5]{Jérôme Chambert}

\author[5]{Aflah Elouneg}

\author[1]{Camilo A. Suarez-Afanador}

\author[1]{Stéphane P. A. Bordas}

\author[3,4]{Giuseppe Sciumè}

\author[2]{Pierre-Yves Rohan*}
\ead{pierre-yves.rohan@ensam.eu}
\cortext[*]{Corresponding author}

\affiliation[1]{organization={Institute of Computational Engineering, Department of Engineering, University of Luxembourg},
addressline={2 place de l'université},
city={Esch-sur-Alzette},
postcode={L-4365},
country={Luxembourg}}

\affiliation[2]{organization={Institut de Biomécanique Humaine Georges Charpak, Arts et Métiers Institute of Technology},
addressline={151 boulevard de l'hôpital},
city={Paris},
postcode={F-75013},
country={France}}

\affiliation[3]{organization={CNRS, Bordeaux INP, I2M, UMR 5295, I2M Bordeaux, Arts et Metiers Institute of Technology, University of Bordeaux},
city={Talence},
postcode={F-33400},
country={France}}

\affiliation[4]{organization={Institut Universitaire de France (IUF)}}

\affiliation[5]{organization={CNRS, Institut FEMTO-ST, Université de Franche-Comté},
city={Besançon},
postcode={F-25000},
country={France}}


\begin{abstract}

This paper proposes a proof of concept application of a biphasic constitutive model to identify the mechanical properties of \textit{in vivo} human skin under extension. Although poromechanics theory has been extensively used to model other soft biological tissues, only a few studies have been published for skin, and most have been limited to \textit{ex vivo} or \textit{in silico} conditions. However, \textit{in vivo} procedures are crucial to determine the subject-specific properties at different body sites. This study focuses on cyclic uni-axial extension of the upper arm skin, using unpublished data collected by Chambert et al. Our analysis shows that a two-layer finite element model allows representing all relevant features of the observed mechanical response to the imposed external loading, which was composed, in this contribution, of four loading-sustaining-unloading cycles. {The Root Mean Square Error (RMSE) between the calibrated model and the measured Force-time response was} $8.84\times 10^{-3}$~\si{\newton}.
Our biphasic model represents a preliminary step toward investigating the mechanical conditions responsible for the onset of injury. It allows for the analysis of changes in Interstitial Fluid (IF) pressure, flow, and osmotic pressure, in addition to the mechanical fields. Future work will focus on the interaction of multiple biochemical factors and the complex network of regulatory signals.
\end{abstract}


\begin{graphicalabstract}
\includegraphics[width=\textwidth]{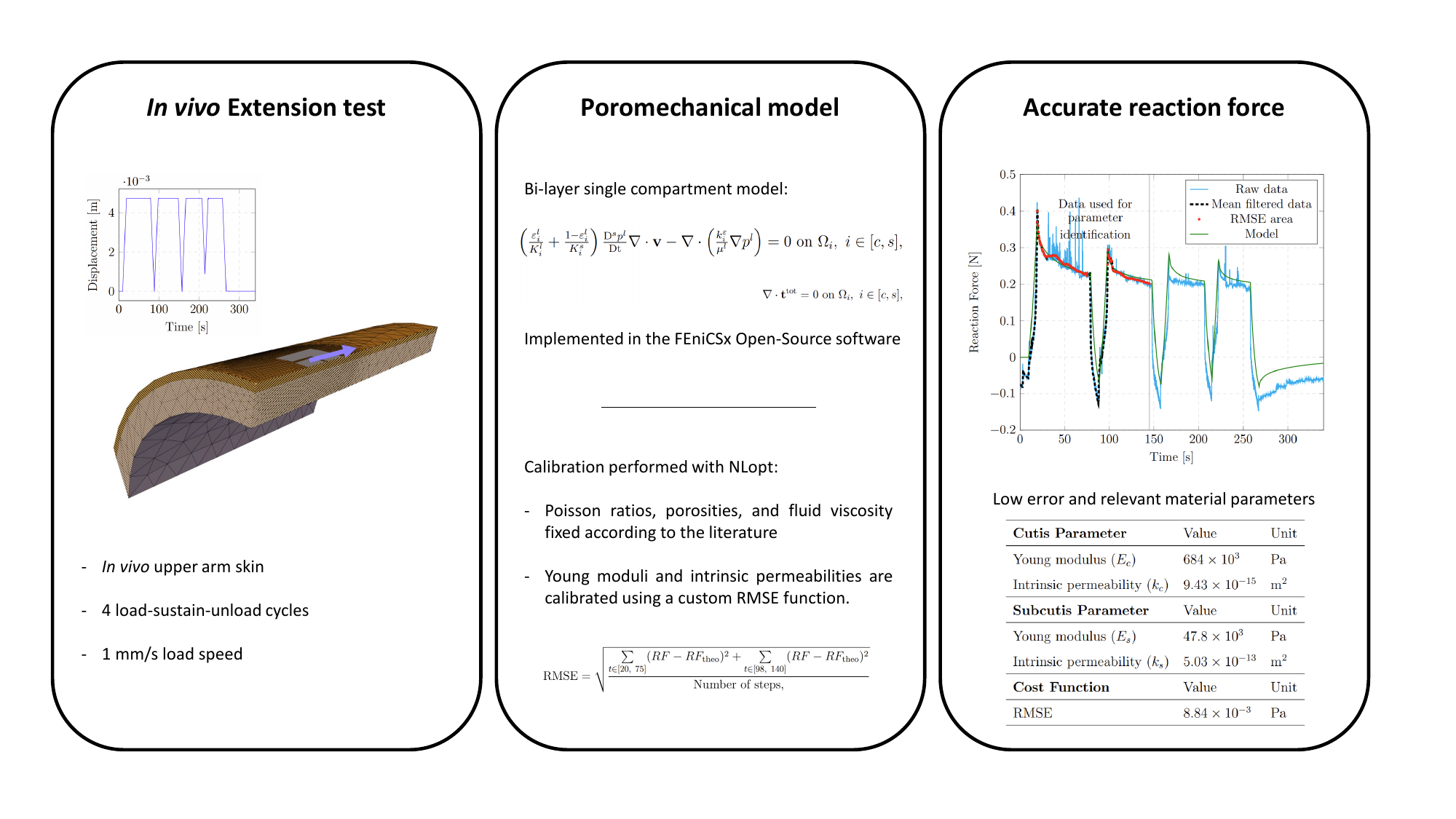}
\end{graphicalabstract}

\begin{highlights}
\item Two-layer poro-hyper-elastic model of the skin
\item \textit{in vivo} experimental data for skin under tension
\item Time-dependent response of the skin during \textit{in vivo} tensile test
\item Implementation of the finite element formulation within FEniCSx
\end{highlights}

\end{frontmatter}

\section{Introduction}\label{sec:introduction}

The skin is a multi-layered structure, the largest organ in the human body. It is vital for protecting the tissues from pollutants, bacterial infections, and sunlight. Knowledge of the ways the skin behaves under mechanical load is crucial for various applications such as skin surgery (\citet{ogawa2012relationship}), diagnostic tools for skin pathology (\citet{jasaitiene2011principles}), design of medical devices or devices for personal care or trans-epidermal drug delivery with micro-needles or micro-jets (\citet{waghule2019microneedles}), and treatment monitoring of skin diseases (\citet{lcoct}). Several \textit{in vivo} measurement tests have been specifically designed to accurately determine the behaviour of skin tissues, such as the suction test (\citet{alexander1977,humbert2017agache}), torsion test (\citet{agache1980mechanical,humbert2017agache}), compressive test (\citet{Oomens1985,Oomens1987,Zhang1994,Pailler-Mattei2004,PaillerMatti2004,Bosboom2001}), indentation test (\citet{PaillerMatti2004}), and longitudinal extension test (\citet{humbert2017agache,Khatyr2004}). 

In particular, these studies considered a broad spectrum of models and have shown the strain rate dependency
of skin mechanical properties (\citet{humbert2017agache,Eshel2001,Shergold2006}). Yet, attempts to characterise the time-dependence of the mechanical response of skin tissue to external loading generally assume a viscoelastic formulation (\citet{Flynn2010,Gerhardt2012}). These models ignore the structural bi-phasic nature of the tissue.

Poroelastic constitutive models have been proposed as an alternative to viscoelastic models to capture the \revision{time-dependent} response of soft tissues. Initially introduced for soil mechanics, these models are largely used in biomechanics. They allow the coupling of the solid behaviour of a scaffold with the fluid mechanics of one or more fluids saturating the solid medium and have been adapted to the biomechanical field(\citet{Argoubi1996,Franceschini2006,Sciume2014,Peyrounette2018,Siddique2017,Gimnich2019,HosseiniFarid2020}). Porous media models represent a promising approach for integrating multiscale/multiphysics data to probe biologically relevant phenomena at smaller scales and embed relevant mechanisms at larger scales. This is particularly the case with the interaction of multiple biochemical factors (enzymes, growth factors, hormones, proteins) and the complex network of regulatory signals, which determine tissue characteristics and their evolution in processes such as growth and remodelling (\citet{Eskandari2015-ih}), ageing, and the onset of injuries such as pressure ulcers (\citet{Sree2019-md, Gefen2022-rc}).

A number of research teams have proposed utilising this approach to characterise and model the mechanical response of various tissues, including the brain (\citet{Budday2019, HosseiniFarid2020, Greiner2021, Urcun2022a, urcun_2023,HervasRaluy2023,CarrascoMantis2023}), the liver (\citet{ricken2019computational}), the meniscus (\citet{Kazemi2013,Uzuner2020, bulle:tel-03652547, Uzuner2022}), and muscle tissue (\citet{Lavigne2022}). In these studies, the biological tissues were modelled as biphasic systems with the behaviour governed by the properties of the porous solid and the fluid occupying the pores. This was based on the balance of mass and momentum conservation equations.

Although the theory of poromechanics has been extensively applied in other soft biological tissues, only a few studies have been published for skin, and these were limited to \textit{in silico} and \textit{ex vivo} studies. 

One of the first contributions was made by (\citet{Oomens1987b}). The authors developed a poroelastic model of the skin based on the large strain theory, using a hyper-elastic (\citet{tong1976stress}) constitutive model for the solid phase and a nonlinear strain-dependent permeability. The model was implemented in a Finite Element model to simulate the quasi-static indentation response of \textit{ex vivo} porcine skin (\citet{OOMENS1987c}). 
More recently, \citet{WeirWeiss2023} developed a custom setup to apply a pressure-driven fluid flow across skin tissues. The resulting flow rate and cross-sectional image acquisition by optical coherence tomography (OCT) combined with digital image correlation (DIC) were used to calculate the internal strains within the tissue and to characterise ex vivo local strain and permeability of porcine skin tissue under compression loading. 
Likewise, \citet{Wahlsten2023-uv} analysed the mechanical behaviour of skin from the cellular to the tissue length scale through dedicated experiments to resolve these discrepancies. Of particular interest, the same team also conducted uni-axial monotonic, cyclic, and relaxation experiments on a total of 37 human and 33 Murine skin samples (\citet{Wahlsten2019}). 
They showed that skin volume is significantly reduced due to tensile elongation. However, the loading was not controlled and applied by a human operator. To the best of the authors' knowledge, little application of biphasic constitutive modelling of the skin \textit{in vivo} has been proposed. Yet, the development of \textit{in vivo} procedures is crucial to determine subject-specific properties at different body sites given different loading conditions.

In this study, we focus on the cyclic uni-axial extension of the upper arm skin from unpublished data previously collected by Chambert et al. under conditions similar to those described in \citet{Chambert2019}. The relaxation phenomenon observed in this experiment is of great interest to the poromechanical approach. Indeed, poromechanics takes into account the \revision{pressurisation} of interstitial fluid in the different layers of the skin, which makes an important contribution to this relaxation during sustained tensile loading.

The experimental data are presented followed by the poromechanical model. The values of the parameters are then discussed based on the physiological and modelling literature. The process of identifying the physical parameters is explained and the numerical response of the model over time is shown. This study is an encouraging step towards the use of poromechanics for \textit{in vivo} skin modelling. The authors hope that it will encourage experimentalists to improve their experimental protocols and measure other quantities (e.g. fluid pressures, fluid flows) further to test the reliability of porous media modelling approaches.

\section{Materials and methods}\label{matmeth}


\subsection{Data Acquisition}
Using an ultra-light, in-house portable extensometer described in \citet{Jacquet2017a}, 
Chambert et al. have carried out uni-axial extension measurements on the left upper arm (dorsal aspect) of a 22-year-old female. \revision{The extensometer is a prototype consisting of two sets of pads: one static and one mobile. Each set includes a central pad, where measurements are taken, and a surrounding 'U-shaped' guarding pad. The guarding pad is designed to minimize disturbances from the surrounding skin in the region of interest by applying the same extension to the neighbouring skin, thereby ensuring uni-axial extension. Further details about the device, along with images of the extensometer, can be found in \citet{Jacquet2017a} and \citet{Chambert2019}.}

\begin{figure}[ht!]
    \centering
    \includegraphics[width=0.4\textwidth,trim={0 0 0 0},clip]{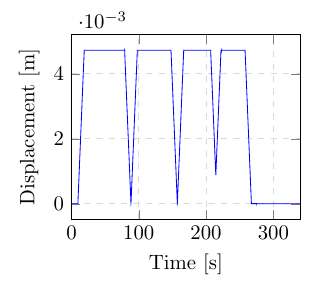}
    \includegraphics[width=0.44\textwidth,trim={0 0 0 0},clip]{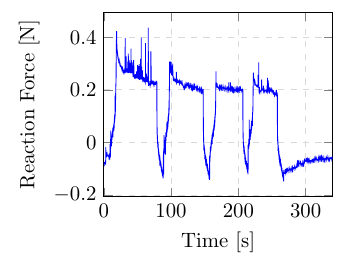}
    \vspace{-1cm}
    \caption{Imposed displacement (half of the displacement extracted from the LVDT sensor for symmetry purpose in the model) and resulting reaction force.} 
    \label{fig:expe}
\end{figure}

During this experimental campaign, four loading-sustaining-unloading repetitions (relaxation tests) were performed at a controlled speed of 1~\si{\milli\meter\per\second} to exhibit the time-dependent properties (Figure~\ref{fig:expe}). The unloading-loading step between the third and fourth cycles was made slightly different, and the real displacement was recorded all along the experiment using a LVDT sensor, providing the exact loading.

The reaction force was monitored every 0.01~\si{\second}, and the maximum applied displacement was set to 9.45~\si{\milli\meter} (Figure~\ref{fig:expe}). This reaction force corresponds to the difference between the current stress state and the pre-stressed state of the skin. \revision{The initial negative reaction force (and the resulting force after unloading) is believed to arise from the installation of the extensometer on the skin. This can be likened to a pre-stress state, which is discussed in the section \ref{sec:disc:skin}.}
All measurements have been made in temperature and hygrometry-regulated rooms (20–22\si{\degree}C, 40–60\si{\percent} relative humidity) after a rest time
of 20~\si{\minute}.

\subsection{Geometrical modelling}

Two curved layers were considered, namely the cutis (composed of the epidermis and the dermis) and the subcutis (defined as the hypodermis) (Figure~\ref{fig:bcs}).
Before the experiment, an echography was performed on the dorsal upper arm using a commercial device (Aixplorer, SuperSonic Imagine, France) with a linear ultrasound probe of 8~MHz central frequency (SuperLinear SL~15-4). \revision{The echography parameters used in this study did not permit the identification of the subcutis boundaries, so the image was utilised solely to measure the thickness of the cutis.} A thickness of approximately 2~\si{\milli\meter} was identified for the cutis\revision{, based on the image scale}. The subcutis thickness was defined based on prior studies (\citet{Birkebaek1998,Mellor2004}). Therefore, a curved bi-layer geometry was considered with a thickness of 2~\si{\milli\meter} for the cutis and 8~\si{\milli\meter} for the subcutis. The radius was fixed to 48~\si{\milli\meter} according to the previous work of \citet{elouneg:tel-04216822}, and based on the measured perimeter of the arm.

Considering the geometry and the boundary conditions (especially of the in-house extensometer geometry and the uni-axial aspect of the experiment), two axes of symmetry were introduced.
The symmetries are represented by the harsh surfaces in Figure~\ref{fig:bcs}. 
A mesh of 11,310 tetrahedral elements was generated using the gmsh software (\citet{gmsh}) and a local refinement close to the in-house device was performed.
Given the strong compatibility between the gmsh and FEniCSx environments, the facets, as well as the elements, were tagged at this stage to apply the boundary conditions. 
The distances on the x- and y- directions were artificially increased to limit the risks of \revision{edge effects} in the region of interest, \textit{i.e.} the region covered by the in-house device.

Another refined mesh of 43,767 elements was also created to assess the trustworthiness of the computed fields. A root mean square error of $8.01\times10^{-3}~\si{\newton}$ was found between the resulting reaction forces obtained by the two meshes. This difference was deemed acceptable given the required increase in computation time which changed from 1\si{\hour}45\si{\minute} to 22\si{\hour}06\si{\minute} with the same configuration of the computer. Hence, the mesh of 11,310 elements was considered for this study.


\subsection{Material modelling}
During the experiment, the skin was stretched up to 25~\si{\percent}. This level of stretching goes beyond the hypothesis of small strains. Furthermore, previous studies highlighted the non-linear behaviour of the skin with its ``J-shaped'' stress-strain curve (\citet{Joodaki2018,MostafaviYazdi2022}). 

A poro-hyper-elastic model was therefore assumed to model the mechanical behaviour of skin tissue. Since the replicated experiment was uni-axial (and supposed to be aligned with the fibre direction of the dermis), for sake of simplicity, the material was assumed to be isotropic, whereas the skin response is known to be transversely isotropic (\citet{Elouneg2023,Kalra2016,MostafaviYazdi2022,Joodaki2018,NAnnaidh2012,Khatyr2004}).

The model is based on the equations of mass and momentum conservation of the fluid and solid phases (Equations~ \eqref{eq:12} and \eqref{eq:13}) as described in \citet{Lavigne2023}. The primary variables in the problem are the pressure in the interstitial space and the solid displacement (assuming the skin is saturated with a single fluid). Considering $\Omega_c$ for the cutis and $\Omega_s$ for the subcutis:

\begin{align}
  \left(\frac{\varepsilon^l_i}{K^l_i}+\frac{1-\varepsilon^l_i}{K^s_i} \right)\frac{\mathrm{D}^s p^l}{\mathrm{D}\text{t}}+\nabla\cdot\mathbf{v} -\nabla\cdot\left(\frac{k^\varepsilon_i}{\mu^l}\mathbf{\nabla}p^l\right) = 0~\text{on }\Omega_i,~i\in[c,s],
  \label{eq:12}\\
 \mathbf{\nabla}\cdot\mathbf{t}^{\text{tot}} = 0~\text{on }\Omega_i,~i\in[c,s],
 \label{eq:13}
\end{align}

\noindent where $\varepsilon^l_i=\frac{\text{Liquid Volume}}{\text{Total Volume}}$ is the porosity of the medium, $K^l_i$ and $K^s_i$ respectively denotes the liquid and solid bulk moduli, $p^l$ is the interstitial pressure, $\mathbf{v}$ is the solid velocity, $k^\varepsilon_i$ corresponds to the intrinsic permeability, $\mu^l$ is the fluid viscosity, and $\mathbf{t}^{\text{tot}}$ is the total stress tensor.

The total stress tensor is expressed as $\mathbf{t}^{\text{tot}}=\mathbf{t}^{\text{eff}}- \beta p^l\mathbf{I}_{\text{d}}$, $\mathbf{t}^{\text{eff}}$ being the effective stress in the sense of porous media mechanics, and assuming that the Biot coefficient ($\beta$) is close to one as typically done in biomechanics (\citet{Lavigne2023}). This assumption results from a bulk modulus of the solid phase $K^s_i$ largely higher than the overall bulk modulus of the porous scaffold (\citet{Scium2021}). To describe the solid behaviour, a Neo-Hookean hyper-elastic potential has been considered because of the straightforward relationship of its parameters (the Lam{\'e} coefficients by $\mu_i=\frac{E_i}{2(1-\nu_i)}$ and $\lambda_i=\frac{E_i\nu_i}{(1+\nu_i)(1-2\nu_i)}$ in $\Omega_i$) with the Young's modulus ($E_i$) and Poisson's ratio ($\nu_i$). This choice was further supported by the ease of interpretation as such energy formulation allows for a direct evaluation of usual elastic parameters. Other strain-energy density functions (a comparison between strain-energy functions for the brain has been proposed in \citet{Budday2019} can be easily introduced in FEniCSx such as performed for the volumetric part in \citet{Lavigne2023}.

Let $\mathbf{F}$ denote the deformation gradient {(Equation~\eqref{eq:21})}, $J$, its determinant and $\mathbf{I}_{\text{d}}$ the identity matrix.The deformation gradient $\mathbf{F}$ reads

 \begin{equation}
   \mathbf{F}=\mathbf{I}_{\text{d}}+\mathbf{\nabla}\mathbf{u},
   \label{eq:21}
 \end{equation}

\noindent where $\mathbf{u}$ is the displacement of the solid phase.

From the deformation gradient $\mathbf{F}$, one can further introduce the right Cauchy-Green stress tensor $\mathbf{C}$ and its first invariant $I_1$ (Equations~\eqref{eq:23} and \eqref{eq:24}). 
 
 \begin{align}
   \mathbf{C} = \mathbf{F}^{\text{T}}\mathbf{F}, \label{eq:23}\\
   I_1 = \text{tr}(\mathbf{C}),
   \label{eq:24}
 \end{align}

\noindent The theory of hyper-elasticity defines a potential of elastic energy $W(\mathbf{F})$ which can be expressed as the combination of an isochoric component and a volumetric component (\citet{SIMO1988,Horgan2004,Michele2018}).

\begin{equation}
  W(\mathbf{F})=\Tilde{W}(I_1,J)+U(J),
  \label{eq:25}
\end{equation}

\noindent where $\Tilde{W}(I_1,J)$ is the isochoric part and $U(J)$ the volumetric one. A compressible formulation of the Neo-Hookean strain-energy potential from (\citet{Pence2014,Horgan2004}) has been introduced as: 

\begin{equation}
  \Tilde{W}(I_1,J)= \frac{\mu_i}{2}(I_1-\text{tr}(\mathbf{I_{\text{d}}})-2\,\ln J),
  \label{eq:26}
\end{equation}

\noindent The volumetric part proposed in \citet{doll2000development} has been adopted: 

\begin{equation}
   U(J)=\frac{K^s_i}{2} (\ln J)^2,
  \label{eq:27}
\end{equation}

\noindent Finally, from the potential (Equation~\eqref{eq:25}) derives the first Piola-Kirchhoff stress tensor. Using Nanson's formula we obtain the effective Cauchy stress such that:

\begin{equation}
  \mathbf{t}^\text{eff}=J^{-1}\frac{\partial W}{\partial \mathbf{F}}\mathbf{F}^T,
  \label{eq:29}
\end{equation}

\noindent A same constitutive law is introduced for the cutis ($\Omega_c$) and the subcutis ($\Omega_s$). To introduce the heterogeneity of the material parameters, they are mapped in the space within the considered finite element software FEniCSx.


\subsection{Boundary and Initial conditions}

\begin{figure}[ht!]
    \centering
    \includegraphics[width=\textwidth,trim={0 0 0 0},clip]{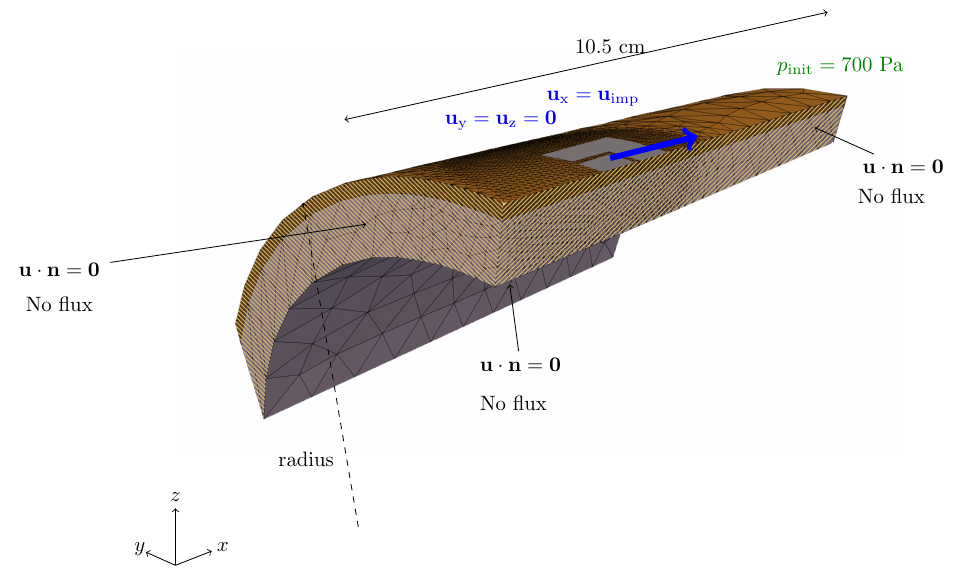}
    \vspace{-1cm}
    \caption{Mesh, boundary and initial conditions of the problem. The skin is assumed saturated by a single fluid. The initial pressure is set to 700~\si{\pascal} to account for the homeostatic pressure. No flux is allowed on the external surfaces. Furthermore, two \revision{planes} of symmetry are introduced \revision{($(x,\,z)$ and $(y,\,z)$)} leading to a normal displacement imposed to zero ($\mathbf{u}\cdot\mathbf{n}=\mathbf{0}$). These are the hashed plans. The vertical displacement is blocked on the bottom surface. The other external surfaces are set free to move. The ``u-shaped'' patch and the patch itself (in gray) have an imposed displacement along the $x$-direction.}
    \label{fig:bcs}
\end{figure}

The poro-hyper-elastic problem is solved using a mixed space (\citet{Lavigne2023}). Therefore, boundary conditions must be introduced for all the primary unknowns, namely the solid displacement of the scaffold and the interstitial pressure (Figure~\ref{fig:bcs}). The mesh not only covers the region of interest but has also been greatly increased to avoid side effects, therefore accounting for a no-flow condition at the boundaries (absence of leakage). No external forces were applied. The amplitude of the displacement applied to the model corresponds to the half of the effective displacement of the experiment recorded by the LVDT sensor. According to \citet{Elouneg2023}, the longitudinal displacement is linearly distributed along the central axis, with the displacement at the centre being half of that imposed on the moving pad. The displacement was applied to the patch and its guide pad (limiting rotational effects resulting from the orientation of the Langer lines in the skin). \revision{It is assumed that the mechanical problem can be addressed using two planes of symmetry $(x,\,z)$ and $(y,\,z)$ (Figure \ref{fig:bcs}), although this approach does not fully capture the actual geometry and loading conditions.} Dirichlet boundary conditions were also introduced on the symmetry planes and the vertical displacement of the bottom surface (Figure~\ref{fig:bcs}). \revision{The two other lateral surfaces were left free to deform to simulate semi-infinite boundary conditions, thereby reducing the risk of edge effects on the computed reaction force. Unlike the symmetry planes, adding Dirichlet boundary conditions was not considered appropriate, as no displacement measurements were taken outside of the extensometer. Imposing such conditions could have interfered with the "skin reorganisation" effect.}

In the absence of experimental data regarding the \textit{in vivo} pre-strain, no pre-stress nor pre-strain was introduced in the model. Regarding initial pressure of the interstitial fluid, experimental literature provides a wide range of \revision{physiological} values: $2144\,\text{Pa}$ for leg's subcutis, averaged from 19 healthy patients, in \citet{hargens1981normal}; from $-670$ to $+536\,\text{Pa}$ for palm hand's subcutis of 7 healthy patients, measured by two techniques, micro-puncture and wick-in-needle, in \citet{wiig1983interstitial}. Additionally, repetitive measurements on pig's skin in \citet{Samant2018}, give the mean osmotic pressure of $725\,\text{Pa}$ (this value is deduced from the increase of initial ionic strength of $4.03448$ fold which provokes an increase of $2200\,\text{Pa}$ of the \revision{homeostatic} pressure). An initial guess of $700\,\text{Pa}$ has been chosen, and the parameter has been included in the sensitivity analysis.


\subsection{Calibration procedure}
\label{sec:calibration}
The poro-hyper-elastic model was implemented in the open-source environment FEniCSx. The implementation process is the same as the one developed in \citet{Lavigne2023}. The corresponding codes are available on \href{https://github.com/Th0masLavigne/Skin_porous_modelling.git}{github} (link \href{https://github.com/Th0masLavigne/Skin_porous_modelling.git}{here}). To limit the effect of local minima during the calibration procedure, only the Young moduli and permeabilities were calibrated for each layer. The other parameters used were fixed according to the literature (see section \ref{sec:initialparam}). To support this choice, the Sobol' indices have been computed and results are reported in \ref{appendix}.

A cost function based on the root mean square error (RMSE) of the reaction force ($RF$) was introduced (Equation~\eqref{eq:rmse}). 
The calibration was only performed in the first 145 seconds to assess the predictive potential of the model. Then, the identified parameters are used to model the full experiment. The cost function was only computed during the time steps of sustained load.

\begin{figure}[ht!]
    \centering
    \includegraphics[width=0.75\textwidth,trim={0 0 0 0},clip]{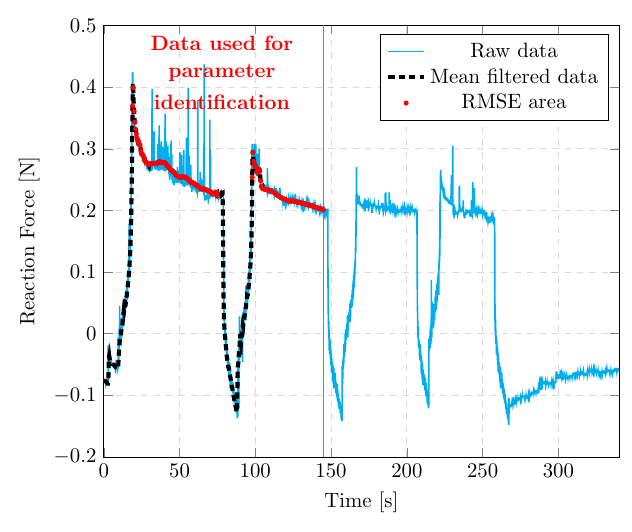}
    \vspace{-1cm}
    \caption{Experimental and filtered curves of reaction force versus time. Due to the presence of noise in the raw data (blue), the calibration area was first filtered (black dashed curve) and the cost function was only evaluated during the sustained displacement (red dots).} 
    \label{fig:expe_calib}
\end{figure}

\begin{equation}
  \text{RMSE}= \sqrt{\frac{\sum\limits_{t\in[20,~75]} (RF - RF_{\text{theo}})^2 + \sum\limits_{t\in[98,~140]} (RF - RF_{\text{theo}})^2 }{\text{Number of steps,}}}
  \label{eq:rmse}
\end{equation}

\noindent A controlled random search (CRS) with local mutation (\citet{Kaelo2006}) from the \href{https://nlopt.readthedocs.io/en/latest/NLopt_Algorithms/}{NLOPT} library was chosen. More specifically, the lower and upper bounds for [$E_c,\,E_s,\,k_c,\,k_s$] were respectively set to [10~\si{\kilo\pascal},\,10~\si{\kilo\pascal},\, $1\times10^{-16}$~\si{\square\meter},\,$1\times10^{-16}$~\si{\square\meter}] and [500~\si{\mega\pascal},\,500~\si{\mega\pascal},\,$1\times10^{-10}$~\si{\square\meter},\,$1\times10^{-10}$~\si{\square\meter}]. Since the parameters' ranges differ from several orders, each parameter was prior normalised with regard to the initial guess to have same order of magnitudes between the optimised unknowns.
The initial population size was set to $3\times(NN+1)$, where $NN$ is the number of parameters ($NN=4$ in the present case). This initial population corresponds to random initial points in the search space differing from the input initial guess (15 in the present case). A stopping criterion of 250 evaluations was added, and the optimal solution was kept. Once the calibration was performed, the set of parameters was applied to reproduce the experiment for its complete duration.


\subsection{Initial parameters values}
\label{sec:initialparam}

Experimental observations and \textit{in silico} calibration techniques (\citet{Khatyr2004,NAnnaidh2012,gallagher2012dynamic,Ottenio2015,Jacquet2017}) have reported a wide range of values for the Young Modulus ranging from 5~\si{\kilo\pascal} to 196~\si{\mega\pascal} (\citet{Payan2017}). The models proposing multi-layer analysis, such as the one proposed by \citet{Han2023}, assume that the cutis is stiffer than the subcutis. This has further been confirmed by the experimental campaigns performed by \citet{Connesson2023} and \citet{PAILLERMATTEI2008599}. As a first guess, the Young's moduli were therefore set to $E_c=1.5\times 10^{6}$~\si{\pascal} and $E_s=5\times 10^{5}$~\si{\pascal}.

Similarly, a wide domain of values has been reported for the permeability of the medium with values ranging from $1\times10^{-16}~\si{\square\meter}$ to $1\times10^{-11}~\si{\square\meter}$ (\citet{Zakaria1997,Swartz2007,Levick1987,Oftadeh2018,deLucio2023,Wahlsten2019}). The initial permeabilities were hence set to $k_c=1\times 10^{{-14}}$~\si{\square\meter} and $k_s=1\times 10^{{-13}}$~\si{\square\meter}.

Concerning the other material parameters, namely the interstitial fluid viscosity, the Poisson ratio, and the porosity, their value was fixed according to the literature. \citet{Sowinski2021} gathered expected viscosity ranges for different human fluids: the cerebrospinal fluid has a viscosity ranging between 0.7-1.0~\si{\milli\pascal\second}, blood between 3.0-67.7~\si{\milli\pascal\second}, ascitic fluid between 0.5-1.5~\si{\milli\pascal\second}.
\revision{\citet{Bera2022} further reported an interstitial fluid dynamic viscosity up to 3.5~\si{\milli\pascal\second}.}
\citet{Swartz2007} respectively reported viscosity values for the plasma, lymph, and synovial fluids of 12~\si{\milli\pascal\second}, 15-22~\si{\milli\pascal\second} and $10^{2}$-$10^{5}$~\si{\milli\pascal\second}. The authors fixed the interstitial fluid viscosity at $\mu^l=5$~\si{\milli\pascal\second}. 
The literature still lacks a clear value of the Poisson ratio, especially for multi-layer models. Therefore, the authors tried to be consistent with values commonly used (\citet{oomens1985mixture,raveh2004elastic,PAILLERMATTEI2008599,Levy2015}) and assumed a solid scaffold with a compressible behaviour for both phases with a higher Poisson ratio value of the cutis of 0.48 and a more compressible subcutis of 0.3. 
Finally, \citet{Samant2018} found that the dermis contains approximately 24\% of \revision{free moving} fluid. The hypodermis is thought to be more porous. The values of the porosities for the cutis and subcutis were respectively fixed at 20\% and 40\%.

The initial guess for the parameters is reported in Table~\ref{tab:param}.

\begin{table}[ht!]
\centering
\begin{tabular}{lll}
\hline
\textbf{Cutis Parameters} & Initial Value & Unit \\ \hline
{Young's} modulus ($E_c$)   &  $1.5\times 10^{5}$ &  \si{\pascal}    \\
\textit{Poisson's ratio}  ($\nu_c$)     &  0.48   &   -   \\
Intrinsic permeability ($k_c$)   & $4\times 10^{{-14}}$ &   \si{\square\meter}   \\
\textit{Initial Porosity}  ($\varepsilon_c^l$)   &   0.2    &   -   \\ \hline
\textbf{Subcutis Parameters} & Value & Unit \\ \hline
{Young's} modulus $E_s$  &  $1\times 10^{5}$  &  \si{\pascal}    \\
\textit{Poisson's ratio} ($\nu_s$)   &  0.3   &   -   \\
Intrinsic permeability ($k_s$)  & $3\times 10^{{-13}}$ &   \si{\square\meter}   \\
\textit{Initial Porosity} {($\varepsilon_s^l$)}   &   0.4    &   -   \\ \hline
\textbf{Fluid phase} & Value & Unit \\ \hline
\textit{IF viscosity} {($\mu^l$)}    &      $5\times 10^{-3}$ &   \si{\pascal\second}   \\ \hline
\end{tabular}
\caption{Initial mechanical parameters for the bi-compartment model. Italic values refer to the fixed parameters during the calibration procedure.}
\label{tab:param}
\end{table}

\section{Results}\label{res__} 

Table~\ref{tab:calib_param} provides the calibrated set of parameters. The minimal cost function value was $8.84\times10^{-3}~\si{\newton}$ (which corresponds to approximately 2\% of the peak reaction force). 

\begin{table}[ht]
\centering
\begin{tabular}{lll}
\hline
\textbf{Cutis Parameter} & Value & Unit \\ \hline
Young modulus   ($E_c$)  &  $684\times10^3$ &  \si{\pascal}    \\
Intrinsic permeability  ($k_c$)   &     $9.43\times 10^{-15}$ &   \si{\square\meter}   \\ \hline
\textbf{Subcutis Parameter} & Value & Unit \\ \hline
Young modulus ($E_s$) &  $47.8\times10^3$  &  \si{\pascal}    \\
Intrinsic permeability ($k_s$)    &   $5.03\times 10^{-13}$   &   \si{\square\meter}   \\ \hline
\textbf{Cost Function} & Value & Unit \\ \hline
RMSE     &      $8.84\times 10^{-3}$ &   \si{\pascal}   \\ \hline
\end{tabular}%
\caption{Calibrated parameters for each layer and optimal cost function value.}
\label{tab:calib_param}
\end{table}

The fluid velocity ($\mathbf{v}^l$) is computed all along the simulation from Darcy's law (Equations (\ref{eq:darcy})) to evaluate the fluid displacement within the tissue during the experiment.
\begin{equation}
    \mathbf{v}^l = -\frac{k}{\mu \varepsilon}\nabla p + \mathbf{v}
    \label{eq:darcy}
\end{equation}

\begin{figure}[ht!]
    \centering
    \includegraphics[width=0.75\textwidth,trim={0 0 0 0},clip]{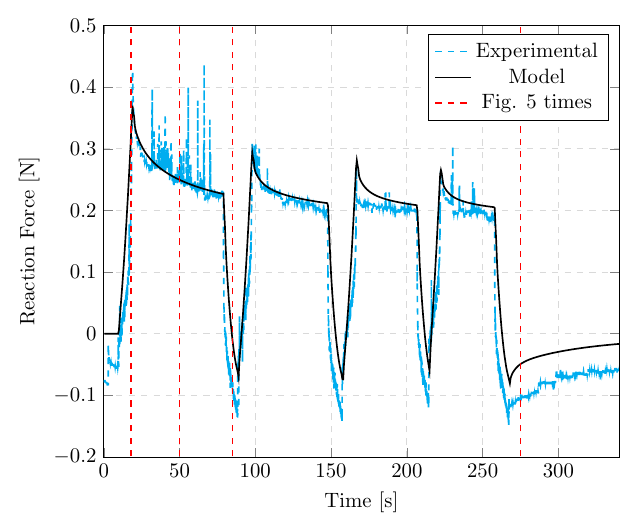}
    \vspace{-1cm}
    \caption{Evaluated response of the model (plain) superimposed with the experimental reaction force (dashed) for all the cycles. The model was able to reproduce the reaction force for the four cycles of the experiment. The vertical dashed lines represent the time steps shown {in} Figure~\ref{fig:timesteps}. }
    \label{fig:calib}
\end{figure}

Figure~\ref{fig:calib} shows the evaluated response on the complete experiment. As reflected by the optimal RMSE value, even if a difference in reaction force for negative values is observed and is constant over time (discussed in the next section), the model response allows reproducing the overall behaviour of the reaction force. Especially, it is worth noting that the model was able to recreate the final reflux even though it was not part of the calibration domain.
It should be noted that the model provides better predictions for the relaxation phases than the loading and unloading phases.

\begin{figure}[H]
    \centering
    \vspace{-4cm}
    \includegraphics[width=\textwidth,trim={0 0 0 0},clip]{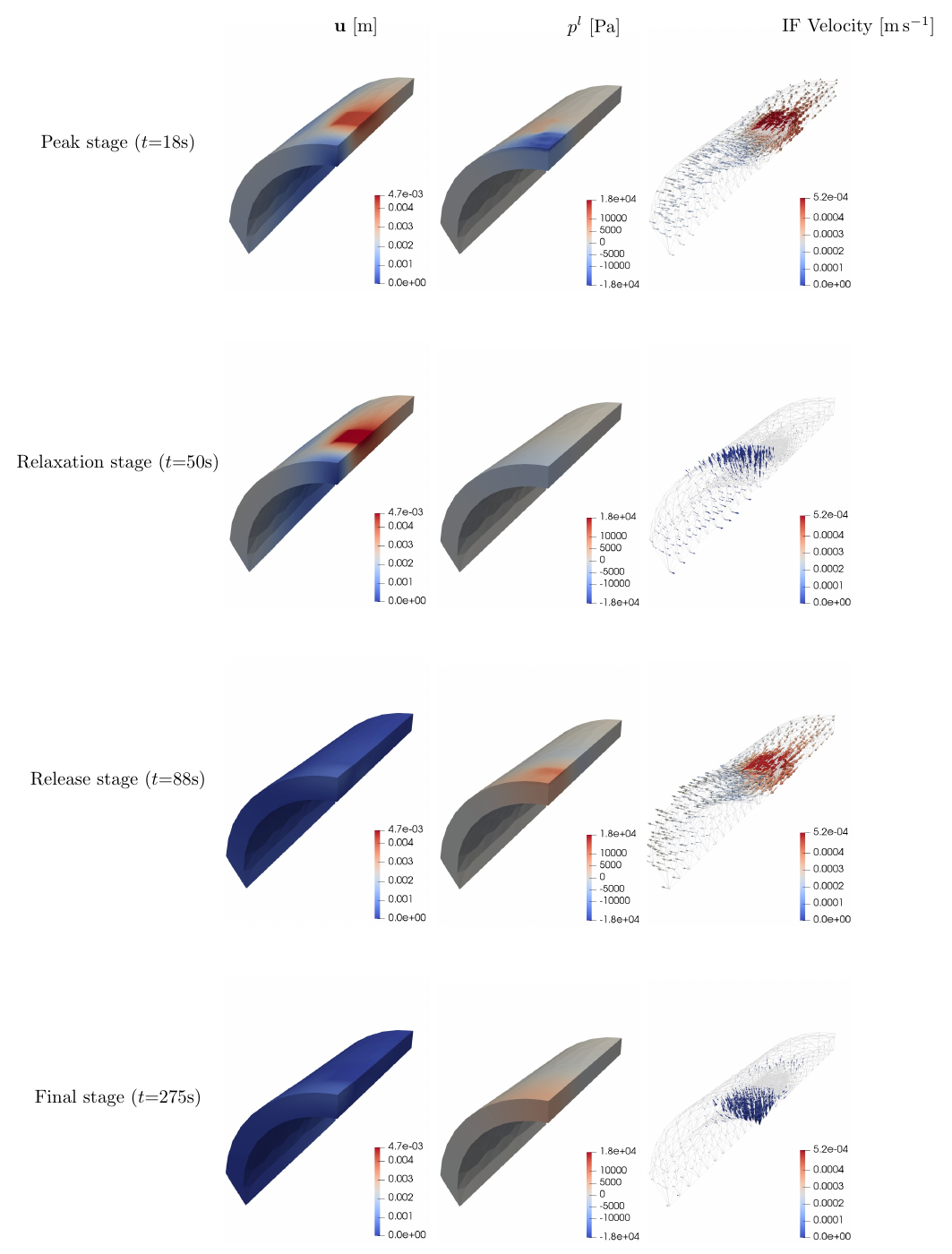}
    \vspace{-1cm}
    \caption{From left to right: displacement magnitude, interstitial fluid pressure, interstitial fluid velocity maps. The results represent the maximum load time ($t=18$~\si{\second}), a sustained phase ($t=50$~\si{\second}), a relaxed phase ($t=88$~\si{\second}) and the final reflux ($t=275$~\si{\second}). When stretching the skin, the fluid tends to be drawn in, resulting in negative values of the interstitial fluid pressure. Conversely, the pores tend to close near to the edge of the u-pad and the fluid is expelled which results in a positive interstitial pressure. During a sustained phase, stabilises towards its homeostatic state. When releasing the tissue, a reflux occurs. The fluid flow images are further included in \textbf{}\ref{appendix:b} for the ease of the reading. A description of the fluid movement is further introduced.}
    \label{fig:timesteps}
\end{figure}

To obtain insight about the physics, Figure~\ref{fig:timesteps} shows the displacement map, the pressure map, and the interstitial fluid flux at different time steps. The authors propose the following interpretation, supporting the time-dependent behaviour of the reaction force from the porous structure.
When stretching the skin, it is assumed that the pores get dilated and, therefore, the fluid tends to be drawn in. 
This was observable on the pressure map through negative values of the interstitial fluid pressure. Conversely, the pores tend to close near the edge of the u-pad and the fluid is expelled which results in a positive interstitial pressure. 

During a sustained phase, the tissue tends to slowly recover its homeostatic state. When releasing the tissue, a strong reflux occurs with high {positive} pressure values: the fluid retrieves its initial place. Finally, the last reflux slowly comes back to a homogenised pressure state within the tissue. 

It is furthermore worth noting that, due to a lower value of the permeability in the cutis phase, most of the fluid flow is concentrated in the subcutis for all the phases.

\clearpage

\section{Discussion}
\label{sec:disc:skin}
Motivated by the limited \textit{in vivo} application of biphasic constitutive modelling for parameter identification, a poro-hyper-elastic model has been implemented in the FEniCSx environment and calibrated to reproduce the time-dependent mechanical response of upper-arm skin under cyclic uni-axial extension including the final reflux phase. The Root Mean Square Error between the calibrated model and the measured Force-time response was $8.84\times 10^{-3}$ \si{\newton}. The calibrated parameters align with previously reported values, confirming that the subcutis is softer than the cutis (section \ref{sec:initialparam}). The {Young's modulus of the cutis ($E_c=684$~\si{\kilo\pascal})} maintains the body geometry, being almost 10 times stiffer than the subcutis. Specifically, the Young's modulus of the cutis ($E_c=684$\si{\kilo\pascal}) functions to absorb external mechanical loads. Interstitial fluid presence contributes to load cushioning and sustains the time-dependent mechanical response. A single fluid compartment was considered for each layer, and permeability was calibrated to reflect interstitial fluid movement. The cutis' intrinsic permeability is $9.43\times10^{-15}~\si{\square\meter}$, while the subcutis permeability is 50 times higher at $5.03\times10^{-13}~\si{\square\meter}$.

However, this study has limitations. The first one is related to the noise from the sensors which were firstly designed for keloids (\citet{Chambert2019}), which are stiffer than healthy skin. Furthermore, \revision{the authors recognise that the current study is based on a single mechanical trial conducted on one subject, serving as an initial investigation to validate the proposed approach. To strengthen the robustness of the findings, future research will involve an expanded sample size and additional trials. Specifically, this patient presented a keloid on her right upper arm and this experimental data was acquired on the other arm to compare the mechanical properties between the keloid-affected area and the healthy skin. This experimental data had not been utilised further and thus provided an opportunity for evaluating the model.} Despite this, the study demonstrates the poro-hyper-elastic model's \textit{in vivo} performance in loading-sustaining-unloading tests, providing valuable insights into the dissipative effect of human skin. It is however difficult to distinguish between viscoelastic and poroelastic dissipation when only limited boundary condition information is known.

Furthermore, a difference is observed for negative values of the reaction force. No pre-stress was introduced and the gap in force is constant for all the minimal values. Overlooking the pre-stress - which may impact the results of a hyper-elastic formulation, the authors however chose not to set to zero the first values to limit misinterpretation of the results.

\revision{Only two layers were considered in the model: the cutis and the subcutis, similar to the approach taken by \citet{Connesson2023}. The cutis, however, can be further subdivided into the epidermis and dermis, with the epidermis primarily composed of cells and the dermis consisting of cells embedded in an extracellular matrix. Given the thickness of each layer and the scale of the geometry under consideration, modelling them as separate layers would significantly increase the number of elements in the mesh, thereby raising the computational cost. This supports the decision to combine them into a single layer. Future studies could explore the subdivision of this layer and investigate its impact on the mechanical response.}

\revision{Additionally, the skin was stretched up to 25~\si{\percent}. As shown in Figure \ref{fig:timesteps}, while the subcutis layer follows the cutis, the deformation is not uniform throughout its depth; the extension test causes deformation that extends into the subcutis layer. Due to a lack of experimental data, no adherence was modelled at the bottom of the subcutis, but this effect might be more pronounced if adherence were considered, potentially increasing shear within the tissue.}

The model assumes isotropy due to limited data along other directions, though skin's fibre nature suggests at least transversely isotropic behaviour (\citet{Khatyr2004,NAnnaidh2012,Kalra2016,Joodaki2018,MostafaviYazdi2022,elouneg:tel-04216822,Elouneg2023}). Future studies should address this and incorporate imaging techniques for better parameter identification. The solid phase could be extended to a viscous transverse hyper-elastic phase to better fit the fast time constants.
Porous media models represent a promising approach for integrating multiscale/multiphysics data to probe biologically relevant phenomena at smaller scales and embed relevant mechanisms at larger scales. This is particularly the case with regard to the interaction of multiple biochemical factors (enzymes, growth factors, hormones, proteins) and the complex network of regulatory signals, which determine tissue characteristics and their evolution in processes such as growth and remodelling (\citet{Eskandari2015-ih}), ageing, and the onset of injuries such as pressure ulcers (\citet{Sree2019-md, Gefen2022-rc}).

The model, tested against the \textit{in vivo} data, shows promise, particularly in understanding tissue necrosis, pressure ulcer onsets, and prevention. Indeed, the aetiology of Pressure Ulcers (PUs) showed that multiple factors lead to the onset of such complications. Especially, a (PU) is assumed to result both from excessive loading and ischaemic damage, which occur at different time and length scales (\citet{Loerakker2011}). Coupling with a second compartment, representative of microvasculature, could offer insights into biochemical and mechanical reactions during mechanical load application (\citet{Scium2013,Kremheller2019,Urcun2021,urcun_2023}).

 The encouraging results obtained in this preliminary work allow for the analysis of changes in IF pressure and flow and osmotic pressure, in addition to the mechanical fields. It represents a first step towards investigating the mechanical conditions responsible for tissue characteristics and their evolution in processes such as growth and remodelling, ageing, and the onset of injuries such as pressure ulcers, this preliminary study focused on applying poromechanical models. More specifically, the proposed modelling approach potentially can pave the way for a better understanding of fibrosis phenomena such as keloid disorders by taking into account the time-dependent mechanical behaviour, which is not the case in recent literature (\citet{Sutula2020,elouneg2021open,elouneg:tel-04216822}).

\section{Conclusion}

This paper aimed to evaluate a poromechanical model to reproduce unpublished data collected on \textit{in vivo} human skin during extension tests. A complete framework to identify the time-dependent properties of the skin was introduced within the Open-Source environment FEniCSx. The calibration was performed on the first cycles of the experiment. The complete experiment was computed afterwards to evaluate the predictability capacity of the model. A good agreement has been found between the numerical and experimental responses. Furthermore, the identified mechanical parameters are relevant with the literature. This study therefore supports the use of a poro-mechanical model for the skin under extension testing. More experiments are however required with, for the IF phase, a better control on the boundary conditions and a better monitoring during the experiment. This would allow to distinguish between viscoelastic and poroelastic dissipation.


\section*{Acknowledgments}
This research was funded in whole, or in part, by the Luxembourg National Research Fund (FNR), grant reference No. 17013182. For the purpose of open access, the author has applied a Creative Commons Attribution 4.0 International (CC BY 4.0) license to any Author Accepted Manuscript version arising from this submission.
The present project is also supported by the National Research Fund, Luxembourg, under grant No. C20/MS/14782078/QuaC and the {French National Research Fund (ANR)} under grant No. ANR-21-CE45-0025 {for the project S-KELOID}. The experimental part of this work has been achieved in the frame of the EIPHI Graduate school (contract "ANR-17-EURE-0002")


\section*{Declarations}
\textbf{Competing interests:} The authors declare that they have no known competing financial interests or personal relationships that could have appeared to influence the work reported in this paper.

\textbf{Supplementary material:} The python codes corresponding to the FEniCSx models and the experimental data of this article are made available on the following link: \url{https://github.com/Th0masLavigne/Skin_porous_modelling.git}.

\appendix

\section{Sobol' indices}\label{appendix}

Sobol' indices were computed for the poromechanical material parameters and the initial IF pressure. More precisely, a 'reference' evaluation was computed and the reaction force was recorded. Each parameter was then moved by 10\% forward and backward, and the resulting reaction force was saved.

The error metric defined to evaluate the impact of the parameter is the mean of the relative difference of the reaction force between 18 s and 50 s. The slope of the curve between the error metrics ($\theta_i$) is then computed for each parameter, and the Sobol' indices are estimated as $S_i=\frac{\theta_i^2}{\sum\theta_i^2}$. The parameters accounting for 90\% of the variance are considered for the calibration.

\begin{table}[ht!]
\centering
\begin{tabular}{lll}
\hline
                & $\theta_i$           & $S_i$                \\ \hline
$E_c$           & $28.7$               & $5.27\times10^{-1}$  \\
$E_s$           & $25.4$               & $4.12\times10^{-1}$  \\
$\varepsilon_c$ & $-2.44\times10^{-5}$ & $3.80\times10^{-13}$ \\
$\varepsilon_s$ & $-2.06\times10^{-4}$ & $2.71\times10^{-11}$ \\
$k_c$           & $-1.6\times10^{-1}$  & $1.64\times10^{-5}$  \\
$k_s$           & $-9.71$              & $6.02\times10^{-2}$  \\
$p_{init}$      & $-1.09\times10^{-4}$ & $7.63\times10^{-12}$ \\ \hline
\end{tabular}%
\caption{Sobol indices of the model's parameters.}
\label{tab:appendix}
\end{table}

According to Table \ref{tab:appendix}, the governing parameters are the Young moduli and the permeability of the subcutis. The permeability of the cutis was added to the calibrated parameters to complete the set of guesses.

\section{Interstitial Fluid flow}\label{appendix:b}
$t=\SI{18}{\second}$
\begin{center}
\includegraphics[trim=0cm  0 0 0cm, clip,width=0.45\textwidth]{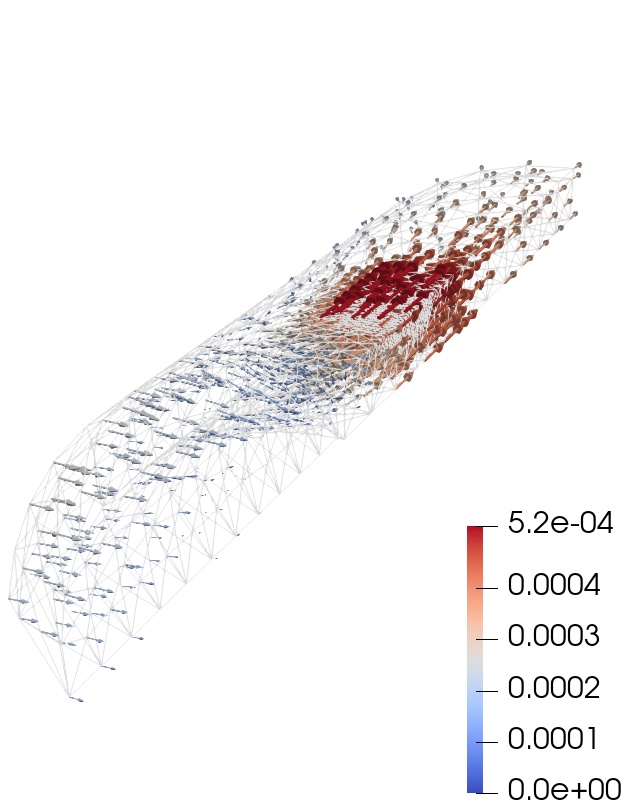} 
\end{center}

$t=\SI{50}{\second}$
\begin{center}
\includegraphics[trim=0cm  0 0 0cm, clip,width=0.45\textwidth]{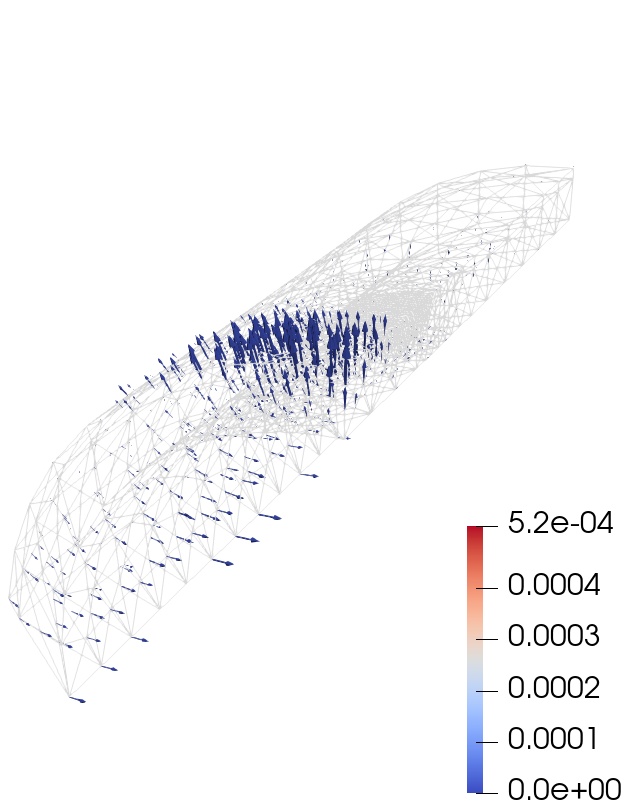}
\end{center}

$t=\SI{88}{\second}$
\begin{center}
\includegraphics[trim=0cm  0 0 0cm, clip,width=0.45\textwidth]{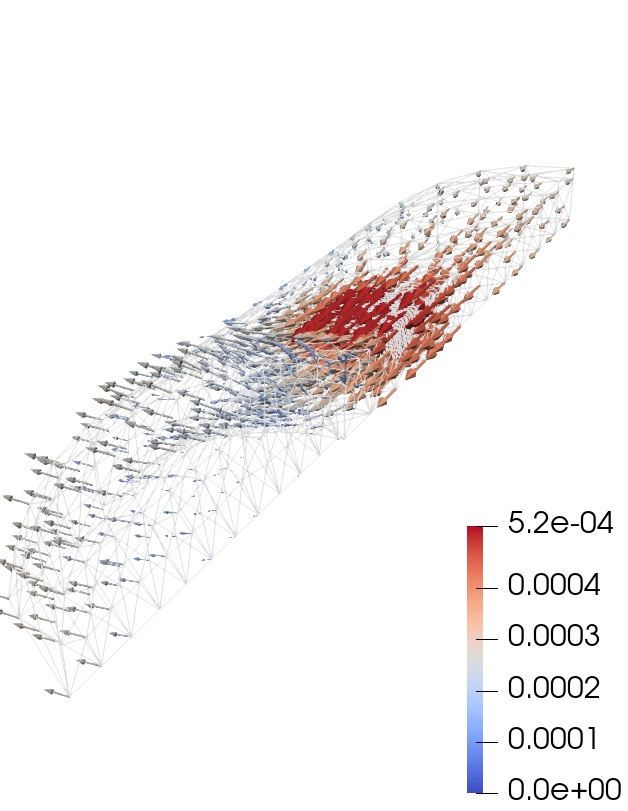}
\end{center}

$t=\SI{275}{\second}$
\begin{center}
\includegraphics[trim=0cm  0 0 0cm, clip,width=0.45\textwidth]{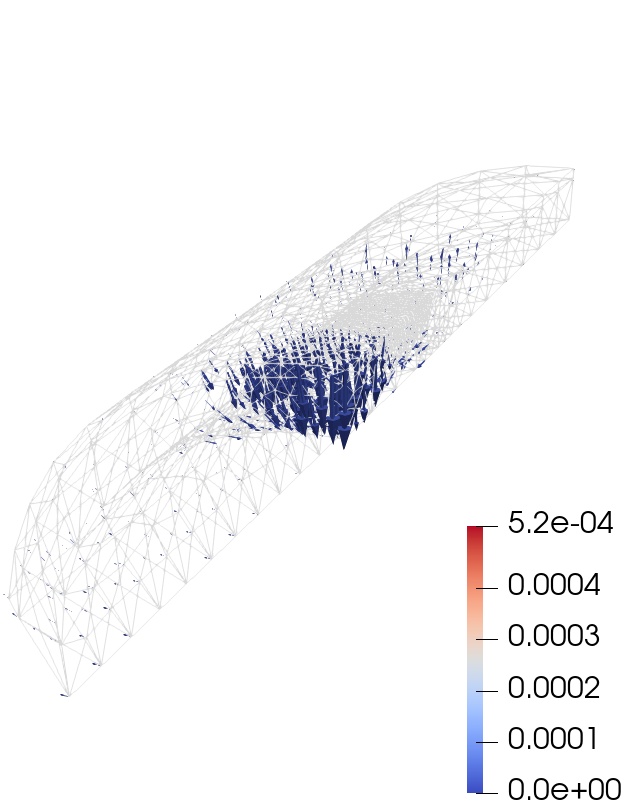}
\end{center}

When stretching the skin, the fluid underneath the patch slowly follows the patch movements and the neighbouring fluid is drawn in the pores which get dilated (peak at t=18s). During a sustained phase (t=50s), an equilibrium phase is reached and the fluid moves from the subcutis to the cutis. Then, during a release phase (t=88s), the fluid is expelled from the deformed region to be replaced by the one underneath the patch. Finally, during the last stage (t=275s), the initial state is reached again and a movement of fluid from the cutis to the subcutis is expected.

\bibliographystyle{apalike} 
\bibliography{references}

\end{document}